**Approaching the ultimate superconducting properties of (Ba,K)Fe$_2$As$_2$ by naturally formed low-angle grain boundary networks**

Kazumasa Iida[1,7,*], Dongyi Qin[2], Chiara Tarantini[3], Takafumi Hatano[1,7], Chao Wang[4], Zimeng Guo[5,7], Hongy Gao[4], Hikaru Saito[6,7], Satoshi Hata[4,5,7], Michio Naito[2,7], Akiyasu Yamamoto[2,7]

[1] Department of Materials Physics, Nagoya University, Furo-cho, Nagoya 464-8603, Japan

[2] Department of Applied Physics, Tokyo University of Agriculture and Technology, Koganei, Tokyo 184-8588, Japan

[3] Applied Superconductivity Center, National High Magnetic Field Laboratory, Florida State University, Tallahassee, United States of America

[4] The Ultramicroscopy Research Center, Kyushu University, 744 Motooka, Nishi, Fukuoka 819-0395, Japan

[5] Interdisciplinary Graduate School of Engineering Sciences, Kyushu University, Kasuga, Fukuoka 816-8580, Japan

[6] Institute for Materials Chemistry and Engineering, Kyushu University, Kasuga, Fukuoka 816-8580, Japan

[7] JST CREST, Kawaguchi, Saitama 332-0012, Japan

*Corresponding author. E-mail: iida@mp.pse.nagoya-u.ac.jp

**Abstract**

The most effective way to enhance the dissipation-free supercurrent in presence of magnetic field for type II superconductors is the introduction of defects that acts as artificial pinning centres (APCs) for the vortices. For instance, the in-field critical current density of doped $BaFe_2As_2$ (Ba122), one of the most technologically important Fe-based superconductors, has been improved over the last decade by APCs created by ion-irradiation. The technique of ion-irradiation has been commonly implemented to determine the ultimate superconducting properties. However, this method is rather complicated and expensive. Here, we report on a surprisingly high critical current density and strong pinning efficiency close to the crystallographic *c*-axis for a K-doped Ba122 epitaxial thin film without APCs, achieving performance comparable to ion-irradiated K-doped Ba122 single crystals. Microstructural analysis reveals that the film is composed of columnar grains having width around 30–60 nm. The grains are rotated around the *b*- (or *a*-) axis by 1.5º and around the *c*-axis by -1º, resulting in the formation of low-angle grain boundary networks. This study demonstrates that the upper limit of in-field properties reached in ion-irradiated K-doped Ba122 is achievable by grain boundary engineering, which is a simple and industrially scalable manner.

**Introduction**

Significant progress on the growth of Fe-based superconductors (FBS) thin films has been achieved over the last decade. As a result, high quality, epitaxial thin films of the technologically important FBS [e.g. Fe(Se,Te), doped $Ae$Fe$_2$As$_2$ ($Ae$: alkaline earth elements) and doped $Ln$FeAsO ($Ln$: lanthanoid elements)] are realised on different kinds of single-crystalline substrates and technical substrates[1-5] except for (Ba,K)Fe$_2$As$_2$ (K-doped Ba122). Realisation of epitaxial K-doped Ba122 has been challenging due to the difficulty in controlling volatile potassium. We have recently succeeded in growing K-doped Ba122 epitaxial thin films on fluoride substrates[6], which gives a great opportunity to investigate their electrical transport properties. Our preliminary study shows that grain boundaries (GBs) are present in K-doped Ba122 despite no sign of weak-link behaviours.

GB with a misorientation angle larger than critical angle $\theta_c$ ~9º becomes a detrimental defect to the critical current for most FBS[3, 7, 8]. On the other hand, GB having a small misorientation angle less than $\theta_c$ does not impede the supercurrent flow. Rather, dislocation arrays in low-angle GB (LAGB) contribute to the flux pinning[3, 7, 8], leading to improvements of the critical current properties of FBS thin films. Indeed, several studies have shown a proof-of-principle of this concept by growing P- and Co-doped Ba122 thin films on technical substrates with oxides buffer layers having a different in-plane spread prepared by ion beam assisted deposition (IBAD)[9, 10]. In both compounds, the larger the texture spread of Ba122 within $\theta_c$, the higher the critical current density $J_c$, typically a few MAcm$^{-2}$ at 4 K. Additionally, $J_c$ for the applied field parallel to the crystallographic $c$-axis ($H \parallel c$) is similar to or even higher than that for $H \parallel ab$[10, 11]. It was later demonstrated that enhanced pinning performance is due to LAGBs acting as flux pinning centres[12].

However, other well-known techniques such as irradiation with protons and heavy-ions produce either isotropic or anisotropic defects (i.e. artificial pinning centres, APCs) significantly enhanced $J_c$ above the value obtained by the aforementioned GB engineering. For instance, SmFeAs(O,F) single crystal with columnar defects produced by heavy-ion irradiation exhibits a high self-field $J_c$ of 18~20 MAcm$^{-2}$ at 5 K, which is about 9~10 times the $J_c$ of the pristine sample[13]. Similarly, Ba$_{0.6}$K$_{0.4}$Fe$_2$As$_2$ single crystals with point defects created by 3 MeV proton irradiation shows a self-field $J_c$ of 11 MAcm$^{-2}$ at 2 K, which is ~4.6 times the $J_c$ of the pristine sample[14]. Quite recently, Ba$_{0.6}$K$_{0.4}$Fe$_2$As$_2$ single crystal irradiated by 320 MeV Au ions shows a very high self-field $J_c$ of over 20 MAcm$^{-2}$ at 2 K[15], corresponding to a 12% of $J_d$~166 MAcm$^{-2}$ [16].

Here, we report on a surprisingly high self-field $J_c$ of 14.4 MAcm$^{-2}$ at 4 K and strong pinning efficiency close to the crystallographic $c$-axis for K-doped Ba122 epitaxial thin film with LAGB networks. The pinning force density $F_p$ for $H \parallel c$ exceeds 200 GNm$^{-3}$ at 4 K and above 6 T, which is at a level comparable to the K-doped Ba122 single crystal with Pb-ions irradiation[17].

## Materials and methods

### Thin film growth

K-doped Ba122 thin films were grown on $CaF_2$(001) at 395°C, a slightly lower temperature than in our previous investigation, by a custom designed molecular beam epitaxy using solid sources of Fe, As, Ba and In-K alloy[6]. Here, we used In-K alloy rather than pure K because of the good controllability of K content in the film as well as for safety issue. $CaF_2$ substrate was fixed on the sample holder using a Ag paste to ensure good thermal conduction. Prior to deposition, the substrate was heated to 600ºC, kept at this temperature for 15 min for thermal cleaning, and subsequently cooled to 395ºC. The compositions of all fluxes except for As was monitored in situ by electron impact emission spectrometry (Ba and Fe) and atomic absorption spectrometry (K). The obtained real-time information was fedback to the personal computer that controls Proportional-Integral-Differential (PID) of the resistive heaters. The As flux was provided constantly during growth. Compared with our previous films, the growth parameters (i.e. deposition temperature and evaporation rate for each flux) have been fully optimised as evidenced in Supplementary fig. S1. Unlike our previous investigation, no impurity phases were observed. Additionally, the average full width at half maximum value of the 103 $\phi$-scan is 1.1º, which is smaller than our previous film[6].

### Microstructural analysis by transmission electron microscopy

Cross-sectional samples were prepared by a focused ion beam. Scanning transmission electron microscopy observation was performed by a TEM (JEOL ARM-200F) operated at an acceleration voltage of 200 kV. TEM-based scanning precession electron diffraction (PED) analysis was performed by a TEM (Thermo Fisher Scientific Tecnai G2 F20 equipped with NanoMEGAS ASTAR system) operated at an acceleration voltage of 200 kV. Details of crystal orientation mapping based on PED are described in Ref. 18. In this PED analysis, the convergence semi-angle of the incident electron beam was 1 mrad, the precession angle was 0.55°. The crystal orientation at each measurement point was determined by matching the PED pattern with template patterns pre-generated from the crystal structural data of K-doped Ba122[19] and $CaF_2$ [20]. $\beta$ and $\gamma$ are defined as the angles between [001]$CaF_2$ and [001]K-doped Ba122, and [100] (or [010]) $CaF_2$ and [100] (or [010]) K-doped Ba122, respectively. Note that $\beta$ and $\gamma$ values in this measurement include uncertainty of ~0.4º, which was estimated from the standard deviation of crystal orientation determination on the $CaF_2$ substrate.

### Electrical transport measurements

A small bridge of 38 μm width and 1 mm length was fabricated by laser cutting. The sample was mounted on a rotator holder in maximum Lorentz force configuration. The angle $\theta$ is measured from the crystallographic *ab*-plane. Current – Voltage (*I–V*) characteristics were measured by a 4-probe method into a commercial physical property measurement system [(PPMS) Quantum

Design]. The upper critical fields $H_{c2}$ were defined as 90% of the normal state resistivity. The irreversibility fields $H_{irr}$ were defined as the intersection between the resistivity traces and the resistivity criterion of $10^{-5}$ mΩcm. An electric field criterion of 1 µV/cm is used to estimate $J_c$.

**Magnetic measurements**

Magnetisation measurements were performed on the rectangular-shaped sample using a superconducting quantum interference device magnetometer [SQUID VSM, (MPMS3) Quantum Design] using. The temperature dependence of susceptibility was measured with a magnetic field of 1 mT applied parallel to the *ab*-plane. Magnetic $J_c$ was determined using the Bean model from the field dependence of magnetisation curves.

**Results**

**Microstructure**

As revealed by structural characterisation using X-ray diffraction, K-doped Ba122 was phase-pure and epitaxially grown on $CaF_2$(001) (Supplementary fig. S1). To evaluate the nanostructure of the grain boundaries, a cross section was observed by scanning transmission electron microscopy (STEM, fig. 1a) and analysed by TEM-based scanning precession electron diffraction (PED). The incident direction of the electron beam is approximately parallel to the [110] direction of the $CaF_2$(001) substrate. An annular dark-field (ADF) image in fig. 1a shows columnar grains growing in the *z* direction, which is more clearly seen in a virtual dark field image of 008 reflection (fig. 1b). The width of columnar grains is 30–60 nm. The epitaxial relationship is revealed as (001)[110]K-doped Ba122 ∥ (001)[100]$CaF_2$ by the PED patterns (fig. 1c and d), which is consistent with the structural characterisation by X-ray diffraction (Supplementary fig. S1). Crystal rotations of K-doped Ba122 around the *b*-axis (equivalent to the *a*-axis) and the *c*-axis were calculated from the crystal orientation data separately and are plotted as two-dimensional maps in fig. 1e and f. For clarity, the crystal rotation angles *β* (around the *b*- and *a*-axis) and *γ* (around the *c*-axis) with respect to $CaF_2$ are shown in fig. 1g and h. As clearly seen in the line profiles (fig. 1i), the average grain rotation around the *b*- (or *a*-) axis is $\Delta\beta_{average} = 1.5°$ and around the *c*-axis is $\Delta\gamma_{average} = -1°$ with respect to the ideal values (i.e. $\Delta\beta = \beta - \beta_{ideal}$, where $\beta_{ideal}$ is 0°, and $\Delta\gamma = \gamma - \gamma_{ideal}$, where $\gamma_{ideal}$ is 45°), resulting in the formation of LAGB networks. As can be seen in Supplementary fig. S2, the [001] of K-doped Ba122 was tilted toward the [0$\bar{1}$0] in our coordinate system. The distribution of *β* over the 2880 points shows that a large fraction locates between 0°–3.5° with a peak of 1.5° (Supplementary fig. S2). For completeness, the distribution of *γ* is also shown in Supplementary fig. S3. This fact reflects the results of angular dependent of $J_c$ measurements, which will be discussed later.

**Resistivity measurements**

$T_{c,90}$, defined as 90% of normal state resistivity, of our K-doped Ba-122 thin film is 35.2 K

(Supplementary fig. S4). The zero-resistivity temperature $T_{c,0}$ is 33 K, corresponding to the onset temperature of the diamagnetic signal measured by the temperature dependence of susceptibility. Therefore, the transition width, defined as $T_{c,90} - T_{c,0}$ is 2.2 K.

To determine the upper critical field $H_{c2}$ and the irreversibility field $H_{irr}$, the temperature dependence of resistivity was measured in field up to 16 T (fig. 2a and b). As increasing applied magnetic fields, a clear shift of $T_c$ to lower temperatures together with a broadening of the superconducting transition is observed for both main crystallographic orientations. The broadening of the transition is more obvious for $H \parallel c$ than $H \parallel ab$, however, such broadening is not so significant compared with $Ln$FeAsO[21] due to the weak thermal fluctuation. It is also worth mentioning that the foot structure in the vicinity of zero resistance arising from the presence of high angle GBs, previously observed in Ref. 22, is not present here. Such foot structure is also due to the poor connectivity. Temperature dependence of the upper critical field $H_{c2}$ and the irreversibility field $H_{irr}$ are summarized in fig. 2c. The slopes of $H_{c2}$ in the field range $0 \leq \mu_0 H \leq 2$ T are -20.1 TK$^{-1}$ and -11.5 TK$^{-1}$ for $H \parallel ab$ and $\parallel c$, respectively. Those values are much higher than the counterpart of single crystal[23]. Another feature is that the slope of $H_{irr}$–line for $H \parallel c$ changes at around 2 T (inset of fig. 2c), which is reminiscent of $RE$Ba$_2$Cu$_3$O$_7$ ($RE$: rare earth elements, $RE$BCO) thin films with the $c$-axis correlated defects[24]. To identify the matching field, $\mu_0 H_{irr}$ is plotted as a function $1 - T/T_{irr}$, where $T_{irr}$ is the irreversibility temperature (fig. 2d). As can be seen, the slope of the $\mu_0 H_{irr}$-line changes from 1.77 to 1.02 at 2 T.

**Pinning potential**

To obtain the activation energy $U_0$ for vortex motion at given fields, linear fits of the Arrhenius plots for the resistivity curves are conducted (figs. 3a and b). Based on the thermally activated flux-flow model[25], the slope of linear fits corresponds to -$U_0$. In fact, on the assumption of the linear temperature dependence, $U(T,H)=U_0(H)(1-T/T_c)$, the following two formulae, ln$\rho(T,H)$=ln$\rho_0(H)$-$U_0(H)/T$ and ln$\rho_0(H)$=ln$\rho_{0f}$+$U_0(H)/T_c$, (derived from $\rho(T,H)= \rho_{0f}$exp[-$U(T,H)/T$]= $\rho_{0f}$exp[-$U_0(H)(1-T/T_c)/T$]) are obtained with $\rho_{0f}$ being the pre-factor. As can be seen in fig. 3c, the activation energy $U_0$ for both $H \parallel c$ and $\parallel ab$ shows the same power law relation $H^{-\alpha}$ in low fields up to 2 T: the exponent α is ~0.05-0.07, which indicates that the single vortex pinning prevails. In this regime, $U_0$ for both directions are 12000~13000 K, whereas the respective values of the Ba$_{0.72}$K$_{0.28}$Fe$_2$As$_2$ single crystal with $T_c$=32 K (i.e. underdoped sample) for $H \parallel ab$ and $\parallel c$ at 1 T are 8500 K and 5000 K[26]. Above 2 T, for $H \parallel ab$, α ~0.5 is consistent with a plastic pinning regime[27]. On the other hand, for $H \parallel c$, α is 0.68, which is located between 0.5 and 1, where the exponent α=1 is the theoretical prediction for collective pinning[28]. It is interesting to note that for high field regime (i.e. 13~16 T) $U_0$ of our film is comparable to the single crystals[26].

The relationship between ln[$\rho_0$] and $U_0$ for both orientations is shown in fig. 3d, where the slope of linear fits corresponds to 1/$T_c$. The respective $T_c$ for $H \parallel c$ and $\parallel ab$ are 35.4 K and 35.5

K, which is close to $T_{c,90}$. This perfect scaling justifies the initial assumption of $U(T,H)=U_0(H)(1-T/T_c)$ in a wide range of temperatures.

**Field dependence of $J_c$ obtained from the transport and magnetisation measurements**

Figure 4a shows the in-field $J_c$ properties for the K-doped Ba122 thin film measured by the *I–V* (or current density *J* – electric field *E*) characteristics at various temperatures. *E–J* curves for $H \parallel c$ are shown in Supplementary fig. S5. At 30 K for both $H \parallel c$ and $\parallel ab$, $J_c$ gradually decreases with increasing fields. However, below 25 K $J_c$ is almost insensitive against applied magnetic fields and a high $J_c$ above $2\times10^5$ A cm$^{-2}$ is maintained on the entire investigated field range. The most striking feature is that $J_c$ for $H \parallel c$ is exceeding that for $H \parallel ab$ with decreasing temperature, opposite to the expected intrinsic behaviour related to the anisotropy of $H_{c2}$. Similar features with inverse anisotropy caused by strong *c*-axis correlated defects were observed before, for instance in Co-doped Ba122[29] and *RE*BCO[24, 30, 31]. These results infer that the strong *c*-axis pinning is active at $T \leq 25$ K. It is worth mentioning that $J_c$ peak for $H \parallel c$ is prominent at high temperatures for *RE*BCO but it strongly suppresses with decreasing temperature[32], which is different from FBS.

To prevent overheating of the contact leads/pads and possible sample damage, the *E-J* characterisation was limited at low fields and temperatures. Hence, for completeness, the field dependence of magnetisation to extract $J_c$ was measured on a rectangular sample cut from the same film used for transport measurements on a wider temperature range (Supplementary fig. S6). $J_c$ calculated from the Bean model is shown in fig. 4b. Except for 28 K, $J_c$ has a weak field dependence, which is consistent with the transport $J_c$. At 4 K, self-field $J_c$ reaches 14.4 MAcm$^{-2}$, corresponding to a ~9% of the depairing current density $J_d$[16]. Temperature dependence of $J_c$ measured by electrical transport measurements follow well the magnetization $J_c$ (fig. 4c), although the electric field criterion $E_c$ of the former is higher than the latter. The data at 30 K slightly deviating from the trend is likely due to the fluctuations close to the $T_c$.

The field dependence of $F_p$ calculated from fig. 4a is summarised in fig. 4d. Because of the presence of strong *c*-axis pinning at $T \leq 25$ K, the maximum $F_p$ is always recorded for $H \parallel c$ within our experimental condition (i.e. up to 16 T).

**Angle dependence of $J_c$ obtained from the transport measurement**

To get better understanding of the pinning efficiency, measurements of the angular dependences of $J_c$ were conducted at various temperatures and field strength (fig. 5). For all fields, $J_c$ peaks around $H \parallel c$ ($\theta=90°$) are weak at 30 K, however, they become intense at $T \leq 25$ K. The peak position of $J_c$ around $H \parallel c$ is away from the *c*-axis by ~4°, indicating that "the correlated defects" are slightly tilted. This is because the columnar grains of K-doped Ba122, which creates LAGBs along the grains, grew unidirectionally at an incline of a few degrees with respect to the substrate normal. To clearly see the effect of correlated defects on $J_c$, $J_c$ anisotropy defined as $J_c/J_c^{ab}$, where $J_c^{ab}$ is $J_c$ at $\theta=180°$, is plotted at the fixed magnetic field (fig. 5e-g). The black dashed lines are positioned at 94º to clearly see the $J_c$ peaks. At 4 T and 30 K, $J_c/J_c^{ab}$ is about 0.5 for $H \parallel c$ (fig. 5e), increasing to ~1.6 at low temperatures. This is a clear indication that the strong pinning around $H \parallel c$ is activated between 30 and 25 K. As increasing applied magnetic fields, a full evolution of the angular dependence of $J_c/J_c^{ab}$ can be observed from a roughly regular behaviour with maximum at 180º for $H \parallel ab$ (e.g. 16 T and 25 K) to an almost isotropic one (e.g. 10 T and 25 K as well as 16 T and 20 K), and finally to a behaviour strongly affected by $c$-axis correlated pinning at the lowest temperatures.

**Discussions**

Through microstructural analyses and electrical transport measurements, "$c$-axis correlated defect" in our K-doped Ba122 thin film is identified as low angle grain boundary (LAGB). On the assumption that the mean distance $d$ of correlated pinning is identical to that of the width of K-doped Ba122 grains (i.e. 30–60 nm), the matching field $B_\phi \sim \phi_0/d^2$ is around 2 T at which a kink of $H_{irr}$ is observed (fig. 2c and d, $\phi_0$ being the flux quantum). As shown in fig. 5, this pinning is strongly temperature dependent, which is due presumably to the crossover between the in-plane coherence length of K-doped Ba122 and the defect size. The correlated GBs pinning and networks improve not only self-field $J_c$ but also in-field $J_c$ for $H$ close to the $c$-axis. Consequently, the anisotropy of $J_c$ is inverted with respect to $H_{c2}$. A similar observation was reported in Ref. 33, where the GBs between columnar grains in MgB$_2$ thin films grown by e-beam evaporation worked as pinning centres.

The tilted growth of K-doped Ba122 is due presumably to the geometrical configuration of the deposition sources together with the deposition without rotating substrates. In our setup, vapour flux arrives at the substrate with an oblique angle. Additionally, adatoms are expected to diffuse relatively slow on the substrate, since the substrate temperature was low compared with the melting temperature of K-doped Ba122 (832°C)[34]. Hence, the shadowing effect[35], which limits the formation of new nuclei during the deposition behind initially formed-nuclei, is pronounced, resulting in the inclined columnar growth.

A pinning force density $F_p$ of 114 GNm$^{-3}$ is recorded even at 15 K and 14–16 T (obtained from the transport measurement) and exceeds 200 GNm$^{-3}$ at 4 K and field above 6 T (the data at 4 K is obtained from the magnetisation measurements in fig. 4b). In fig. 6, the field dependence of $F_p$ for our K-doped Ba122 thin film is plotted. For comparison, we also plotted the following data of pinning enhanced Ba122 single crystals and thin films with different dopants: K-doped Ba122 single crystals with Pb-ions irradiation measured at 5 K[17], Co-doped Ba122 thin film with large amounts of stacking faults measured at 4.2 K[35], Co-doped Ba122 thin film with 3mol% BaZrO$_3$ (BZO) measured at 5 K[36], and P-doped Ba122 with 3mol% BZO measured at 4.2 K and 15 K[37]. As can be seen, up to 4 T, $F_p$ of our K-doped Ba122 thin film is the highest among the pinning enhanced Ba122. Albeit the $F_p$ data above 7 T are missing, the extrapolated value at 9 T for the K-doped Ba122 thin film is comparable to the highest value reported for Co-doped Ba122 thin film.

Huge improvement of the superconducting properties of our K-doped Ba122 thin film without APCs is due to a high density of correlated pinning centres created by LAGB networks. Unlike Co- and P-doped Ba122 thin films, the growth temperature of K-doped Ba122 thin films is quite low (~400°C). This low temperature synthesis may lead to small grain size, and hence, the increase of the density of LAGB. It is worth mentioning that the dislocation density increases with increasing the grain boundary angle. Hence, further improvement of in-field $J_c$ is possible by enlarging the texture spread within the critical angle $\theta_c$. Grain boundary engineering present in this study highlights the possible novel approach to improve the superconducting properties, which is a simple and industrially scalable manner.

**Conclusion**

Herein, we have investigated nanoscale microstructure of K-doped Ba122 epitaxial thin film grown on CaF$_2$ by molecular beam epitaxy. The nanoscale crystal orientation mapping shows that the film is composed of columnar grains having width around 30–60 nm. The average grain rotation around the *b*- (or *a*-) axis is 1.5º and around the *c*-axis is -1º with respect to the ideal values, resulting in the formation of low-angle grain boundary networks. Thanks to LAGB networks, superior superconducting properties of K-doped Ba122 are realised: the pinning force density $F_p$ for $H \parallel c$ exceeds over 200 GNm$^{-3}$ at 4 K and above 6 T, which is comparable to the best performing K-doped Ba122 by ion-irradiation.

superconducting films. *Supercond. Sci. Technol*. **32**, 064005 (2019).

**Acknowledgements**

We thank Wai-Kwong Kwok (Argonne National Laboratory) for data[17], Yanwei Ma (Chinese Academy of Science) for data[36], Jongmin Lee and Sanghan Lee (Gwangju Institute of Science and Technology) for data[37], and Masashi Miura (Seikei University) for data[38]. This work was supported by JST CREST Grant Number JPMJCR18J4. A portion of work was performed at the National High Magnetic Field Laboratory, which was supported by National Science Foundation Cooperative Agreement No. DMR-1644779 and the State of Florida. It was also supported by US Department of Energy Office of High Energy Physics under the grant number DE-SC0018750. This work was also partly supported by Advanced Characterization Platform of the Nanotechnology Platform Japan sponsored by the Ministry of Education, Culture, Sports, Science and Technology (MEXT), Japan.

**Author Contributions**

K.I. and A.Y. designed the study. K.I. and C.T. wrote manuscript together with D.Q., H.S., S.H., M.N. and A.Y.. Thin films preparation, structural characterisations by XRD, and micro bridge fabrications were carried out by D.Q., M.N, K.I., T.H. and C.T.. Microstrutural characterisations by TEM were performed by C.W., Z.G., H.G., H.S. and S.H., and C.T. conducted in-field electrical transport measurements.

**Conflict of Interests**

The authors declare that they have no conflict of interest.

**Figures captions**

**Figure 1 | Microstructural analyses by TEM. a,** Cross-sectional view obtained by ADF-STEM. **b,** Virtual dark field image of 008 reflection of K-doped Ba122. **c,** Typical PED patterns extracted from the K-doped Ba122 thin film (red cross in **b**) and **d,** the $CaF_2$ substrate (green cross in **b**). **e,** $\beta$ rotation map and **f,** $\gamma$ rotation map obtained from the K-doped Ba122 thin film. The z-axis shows the distance from the interface between K-doped Ba122 and $CaF_2$, the same direction as shown in **a**. The ideal angles, 0º and 45º, are defined as light-green and red colour in **e** and **f**, respectively. **g,** Schematic illustration the crystal rotation angles $\beta$ (around the [100]-axis) and **h,** $\gamma$ (around [001]-axis) with respect to $CaF_2$ as the reference. **i,** Line profiles of $\beta$ rotation and $\gamma$ rotation extracted along the black broken lines in **e** and **f**, respectively. The lines of $\beta = 0°$ and $\gamma = 45°$ are marked for comparison.

**Figure 2 | In-field resistivity measurements and the magnetic phase diagram for the K-doped Ba122 thin film. a,** Resistivity curves for $H$ parallel to the crystallographic c-axis and **b,** $H \parallel ab$-plane. Field increment was 2 T from 2 to 16 T. Below 2 T, measurements were conducted at 0, 0.5, 1, and 2 T. **c,** Both $H_{c2}$ and $H_{irr}$ are plotted as a function of temperature. The solid symbols represent $H_{c2}$ and the open symbols show $H_{irr}$, respectively. For $H \parallel c$, the slope of $H_{irr}$ changes around 2 T as indicated by the arrow. **d,** Logarithmic presentation of $H_{irr}$ vs $1-T/T_{irr}$, where $T_{irr}$ is the irreversibility temperature at self-field. The slope changes at 2 T, corresponding to the matching field.

**Figure 3 | Arrhenius plots of the resistivity curves shown in figs. 2a and b, and the resultants pinning potential $U_0$ and prefactor $\rho_{0f}$ for the K-doped Ba122 thin film. a,** For $H$ parallel to the crystallographic c-axis and **b,** ab-plane. **c,** Field dependence of the pinning potential $U_0(H)$ for both main crystallographic orientations. **d,** $U_0(H)$ dependent of $\ln[\rho_0 \text{ (m}\Omega\text{cm)}]$ for $H \parallel c$ and $\parallel ab$.

**Figure 4 | Field dependence of the critical current density $J_c$ measured by the transport and magnetisation methods. a,** $J_c$-$H$ characteristics for both main orientations obtained from the transport measurements. The solid symbols represent $H \parallel c$ and the open symbols show $H \parallel ab$, respectively. **b,** Field dependence of $J_c$ evaluated from the magnetisation measurements using the extended Bean model. **c,** Temperature dependence of $J_c$ for several applied fields $H \parallel c$. **d,** Field dependence of $F_p$ calculated from a.

**Figure 5 | Angular dependence of transport $J_c$ for the K-doped Ba122 thin film.** Measurement temperature was **a,** 30 K, **b,** 25 K, **c,** 20 K, and **d,** 15 K. At angles of $\theta$=90º and

180º correspond to $H \parallel c$ and $\parallel ab$, respectively. Angle dependence of $J_c$ normalised by $J_c$ for $H \parallel ab$ ($J_c^{ab}$) measured at **e**, 4 T, **f**, 10 T, and **g**, 16 T, respectively. The dashed lines are located at $\theta$=94º to clearly see the $J_c$ peaks.

**Figure 6 | Field dependence of the pinning force density $F_p$ for $H \parallel c$.** The $F_p$ for K-doped Ba122 thin film measured at 4 K and 15 K. The respective $F_p$ at 4 K and 15 K are calculated using the $J_c$-$H$ data obtained from the magnetic and transport measurements. For comparison, the data of K-doped Ba122 single crystal with Pb-ions irradiation measured at 5 K[17], Co-doped Ba122 thin film with a large amount of stacking faults measured at 4.2 K[36], Co-doped Ba122 thin film with 3 mol% $BaZrO_3$ (BZO) measured at 4.2 K[37], and P-doped Ba122 thin film with 3mol% BZO measured at 4 K and 15 K[38] are also plotted.

**Figure 1**

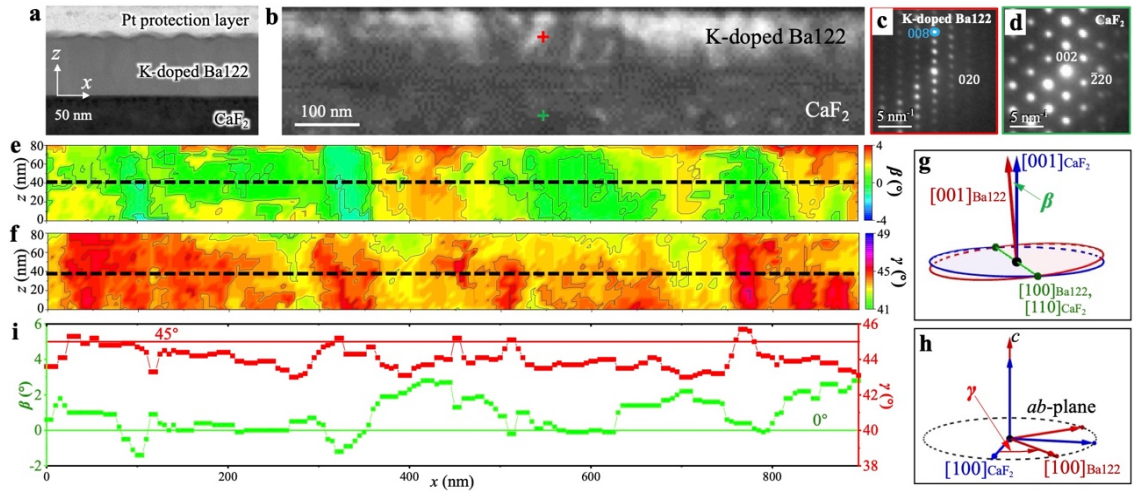

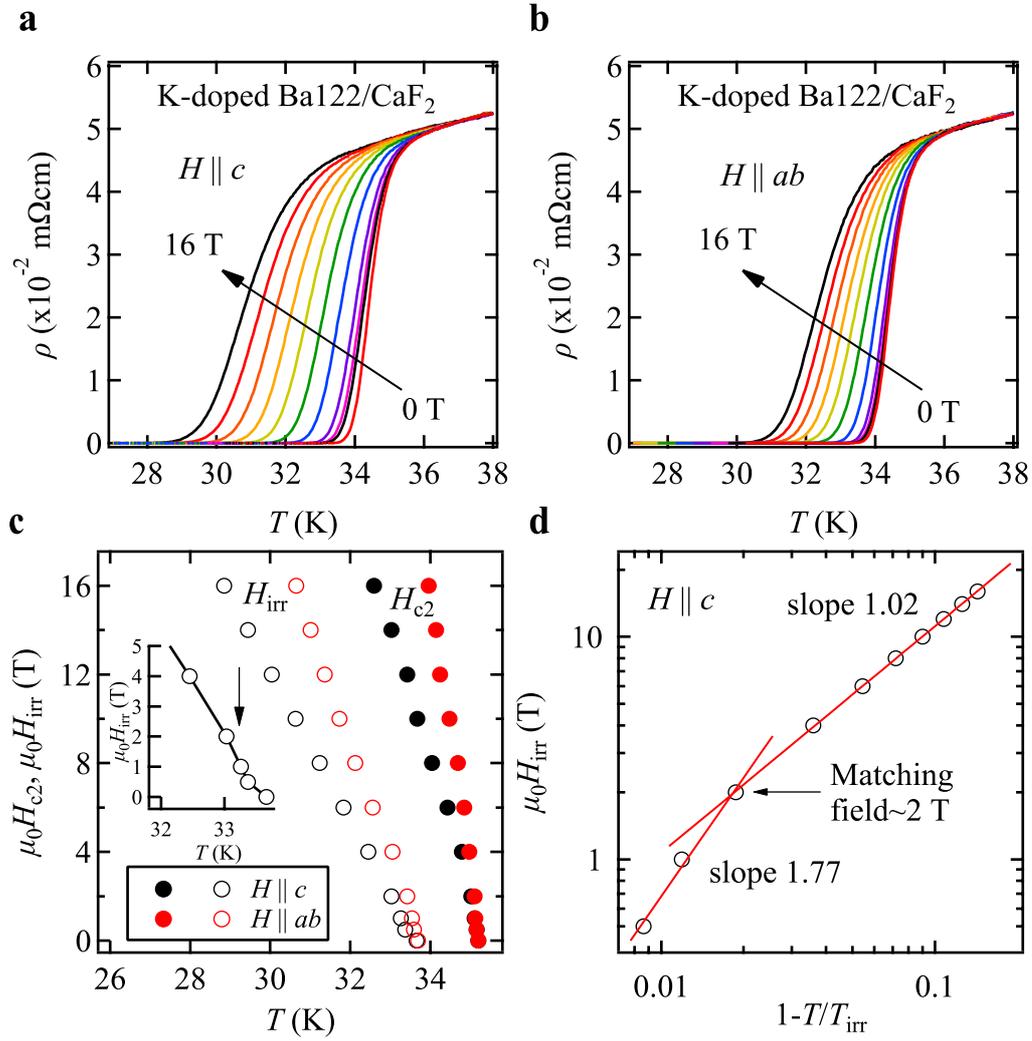

**Figure 3**

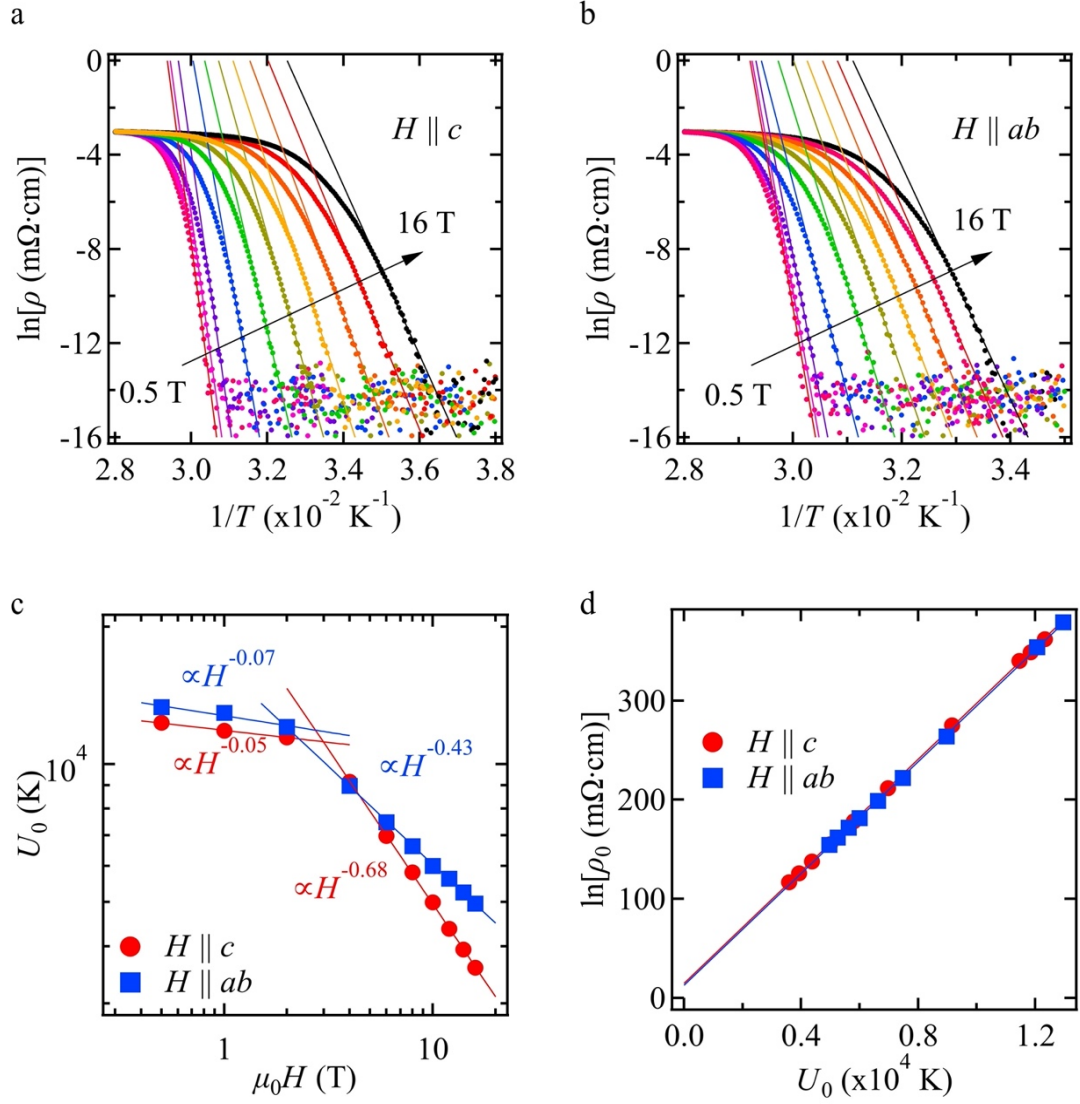

Figure 4

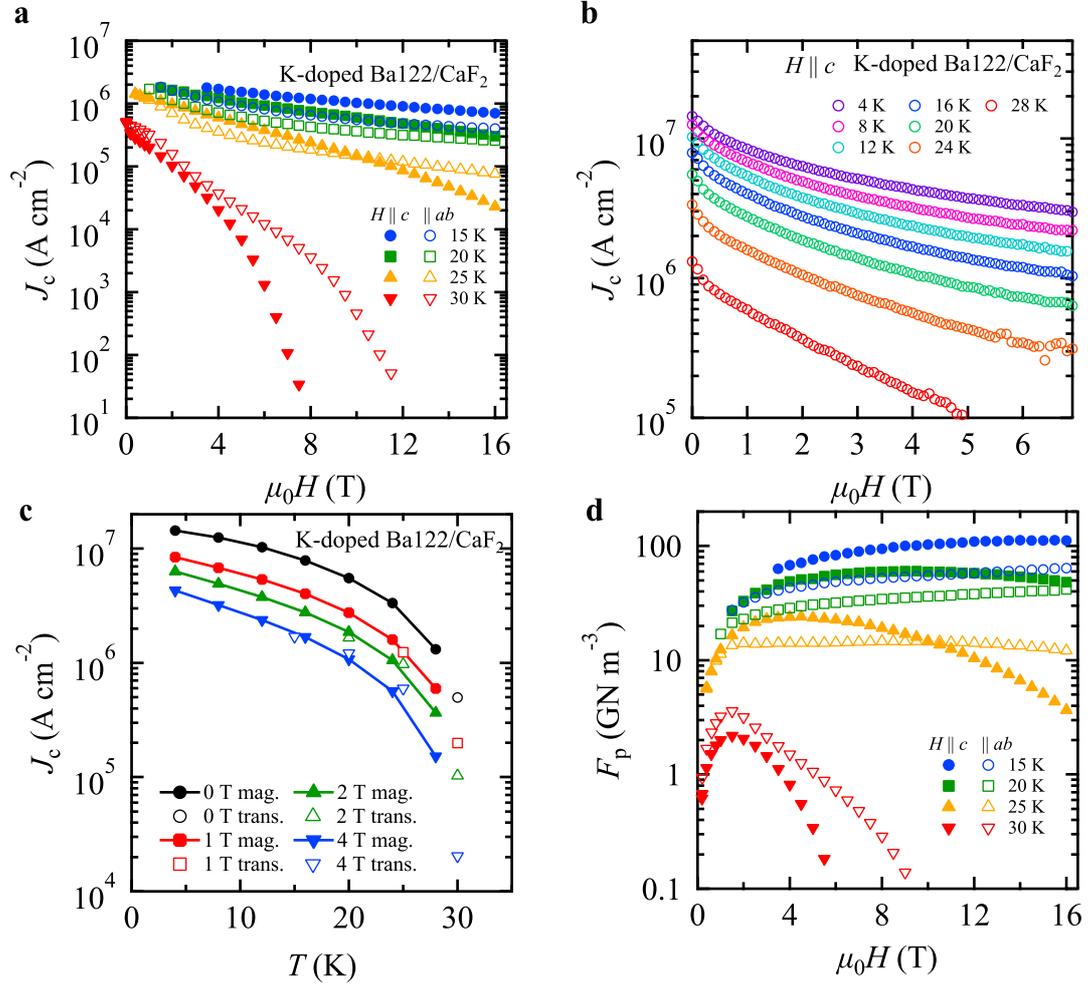

**Figure 5**

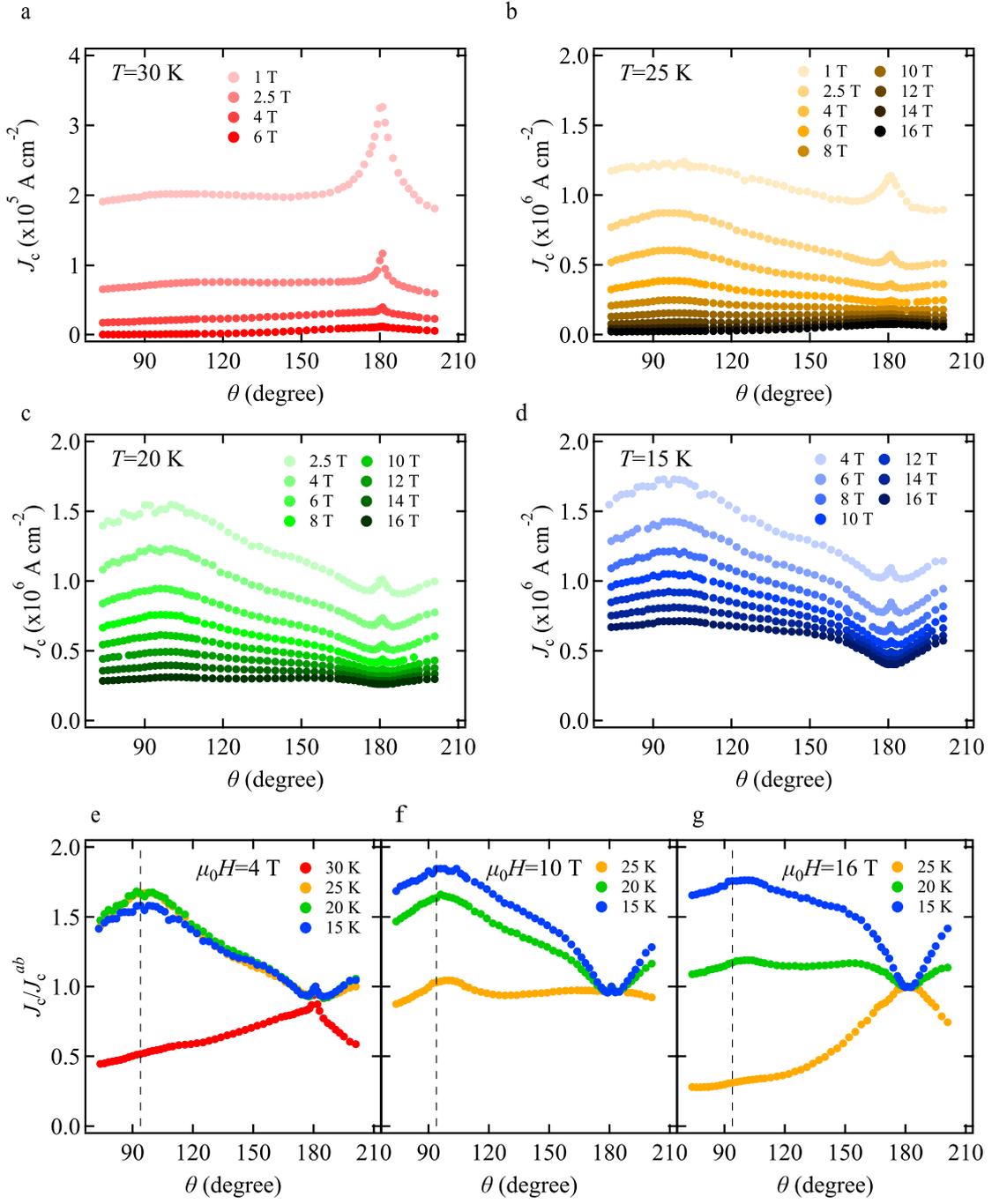

Figure 6

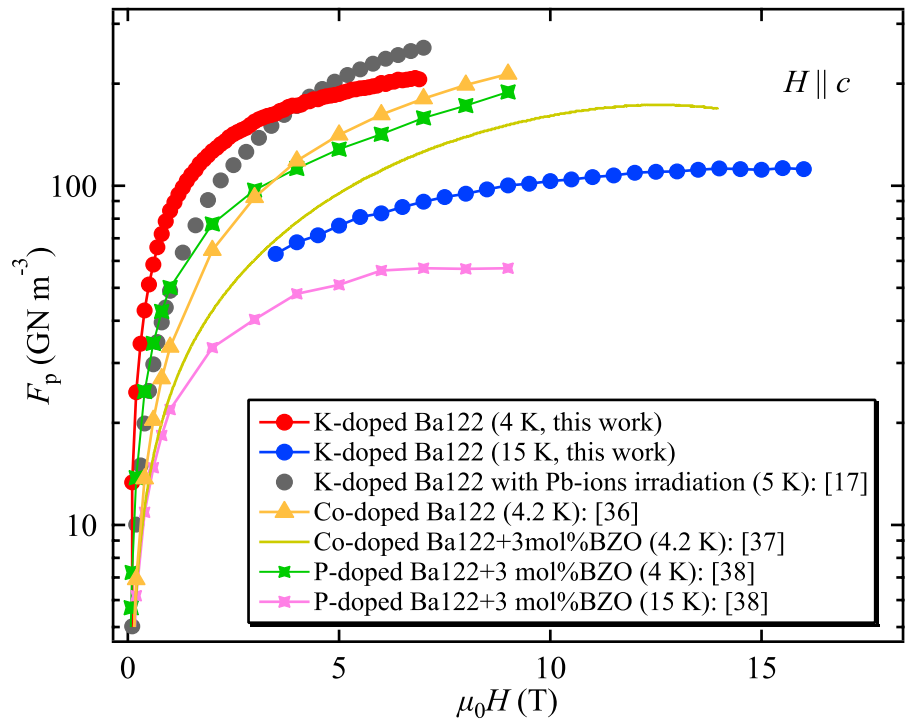

**Supplementary information on "Approaching the ultimate superconducting properties of (Ba,K)Fe₂As₂ by naturally formed low-angle grain boundary networks"**

Kazumasa Iida[1,7], Dongyi Qin[2], Chiara Tarantini[3], Takafumi Hatano[1,7], Chao Wang[4], Zimeng Guo[5], Hongy Gao[4], Hikaru Saito[6,7], Satoshi Hata[4,5,7], Michio Naito[2,7], Akiyasu Yamamoto[2,7]

1 Department of Materials Physics, Nagoya University, Furo-cho, Nagoya 464-8603, Japan
2 Department of Applied Physics, Tokyo University of Agriculture and Technology, Koganei, Tokyo 184-8588, Japan
3 Applied Superconductivity Center, National High Magnetic Field Laboratory, Florida State University, Tallahassee, United States of America
4 The Ultramicroscopy Research Center, Kyushu University, 744 Motooka, Nishi, Fukuoka 819-0395, Japan
5 Interdisciplinary Graduate School of Engineering Sciences, Kyushu University, Kasuga, Fukuoka 816-8580, Japan
6 Institute for Materials Chemistry and Engineering, Kyushu University, Kasuga, Fukuoka 816-8580, Japan
7 JST CREST, Kawaguchi, Saitama 332-0012, Japan

1. **Structural characterisation by X-ray diffraction**

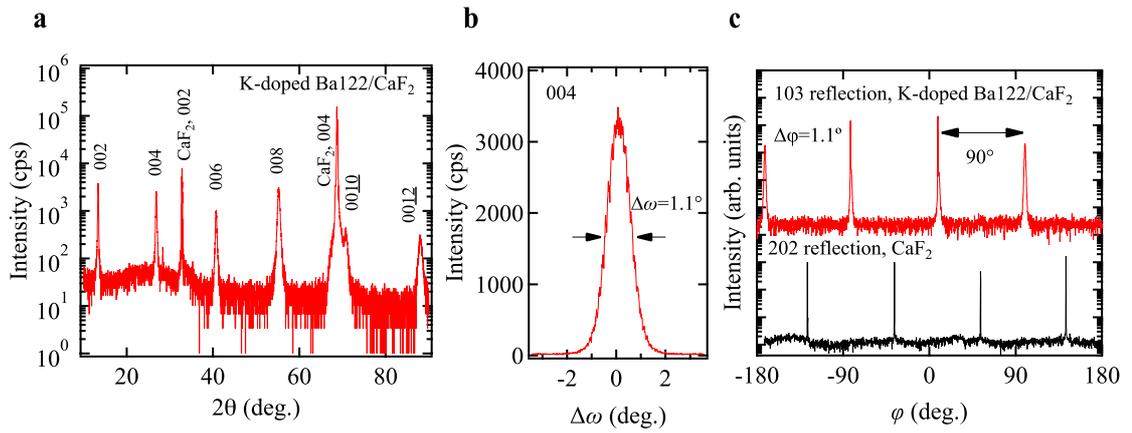

**Figure S1 | Structural characterisation by X-ray diffraction (XRD) for the K-doped Ba122 thin film. a,** $2\theta/\omega$–scan. **b,** Rocking curve for the 004 reflection. **c,** Azimuthal $\phi$-scan of the off-axis 103 reflection of K-doped Ba122 and 202 reflection of CaF₂. It is clear that phase-pure K-doped Ba122 is grown epitaxially on CaF₂. The epitaxial relation is identified as (001)[110]K-doped Ba122 ∥ (001)[100]CaF₂.

## 2. The stereographic projection of K-doped Ba122 on the view direction of [001]CaF$_2$.

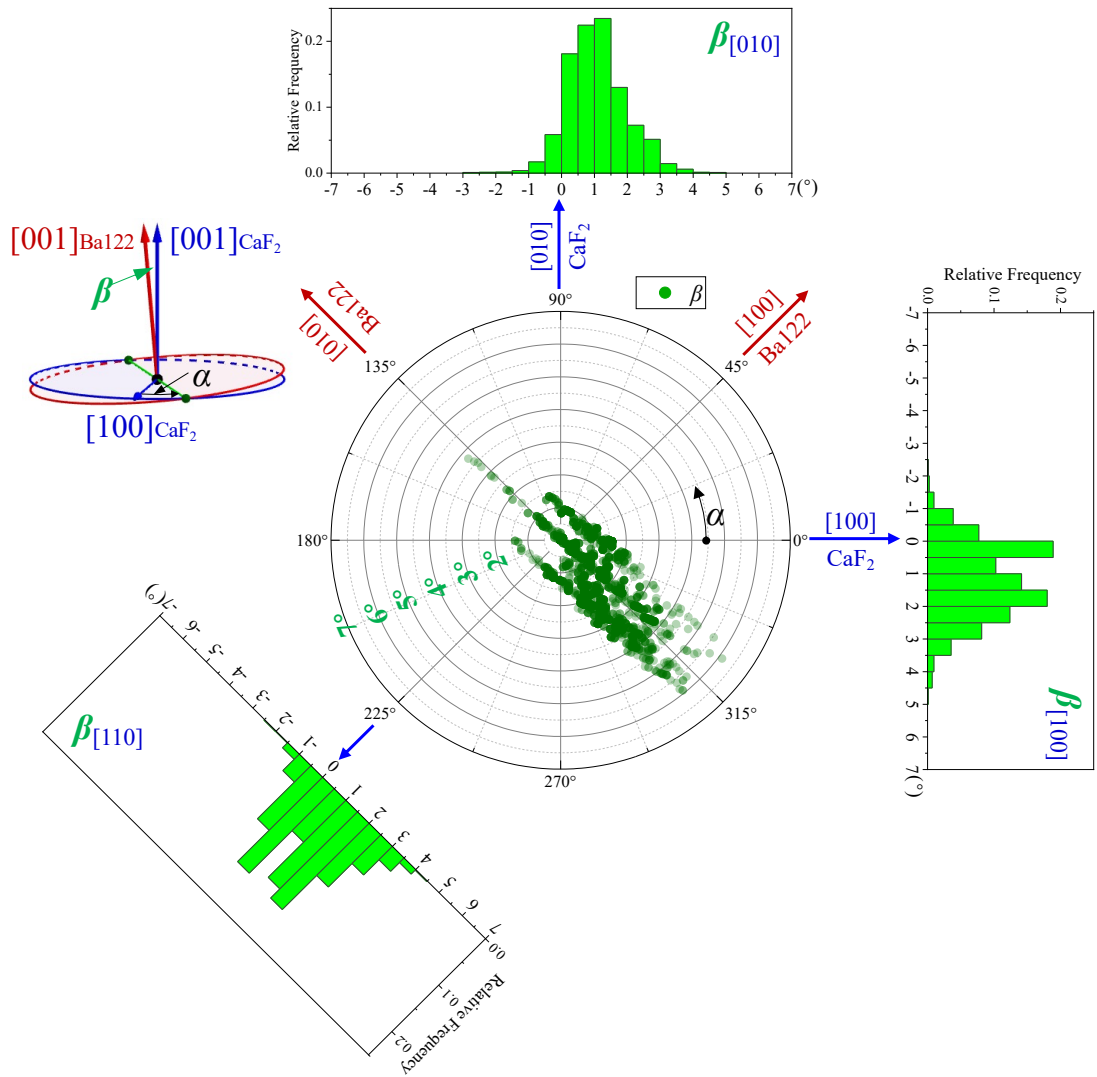

**Figure S2 | The polar figure and the surrounding histograms of the out-of-plane misorientation of the K-doped Ba122 thin film.** The histograms exhibit the distributions of $\beta$ angle on three specific projected direction, [100]CaF$_2$, [010]CaF$_2$ and [110]CaF$_2$ (almost equivalent to the [100] direction of K-doped Ba122). Euler angle $\alpha$ is defined as shown in the inset. As can be seen from this figure, the [001] direction of K-doped Ba122 is tilted toward the [0$\bar{1}$0] and its distribution (i.e. $\beta$) over the 2880 points shows that a large fraction locates between 0°– 3.5° and the peak position lies at around 1.5°.

## 3. In-plane misorientation angles of K-doped Ba122

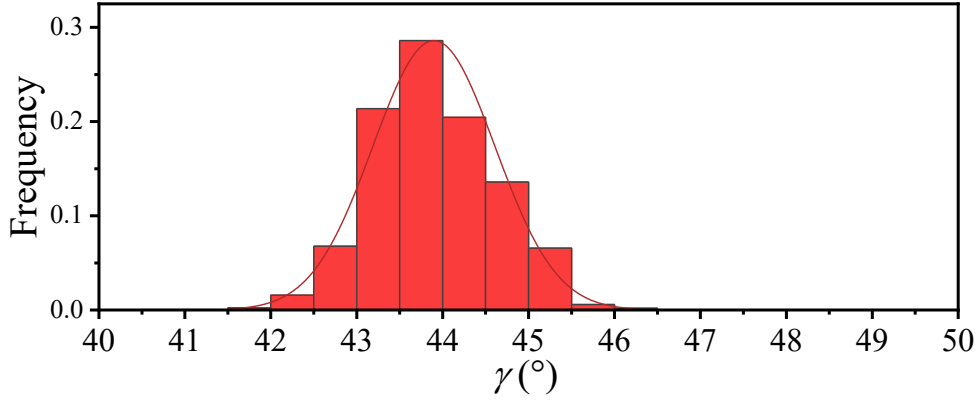

**Figure S3 | The histogram of the in-plane misorientation of the K-doped Ba122 thin film.** The distribution of γ shows that a large fraction lies 43°–45°.

## 4. Superconducting transition temperature $T_c$ determined by resistive and magnetic measurements

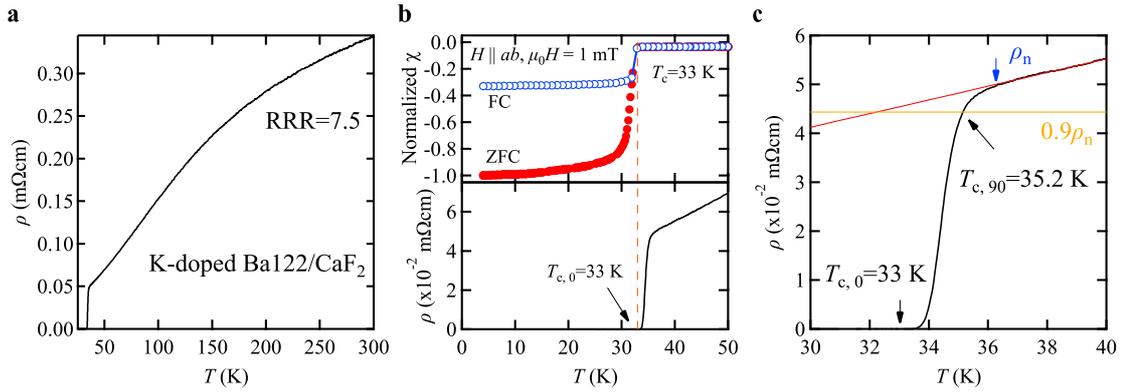

**Figure S4 | Temperature dependence of the resistivity and magnetisation measurements for the K-doped Ba122 thin film. a,** Resistivity curve $\rho(T)$ in the absence of magnetic field. Residual resistivity ratio (RRR), defined as $\rho(300\ K)/\rho_n$ ($\rho_n$: normal state resistivity, see below), is 7.5. Large RRR, low resistivity at 300 K and a sharp superconducting transition of 2.2 K ($T_{c,90} - T_{c,0}$, $T_{c,90}$: Temperature reaches 90% of $\rho_n$, $T_{c,0}$: Temperature reaches zero resistivity) compared to the polycrystalline films [S1] indicate that our K-doped Ba122 film is free of current-blocking impurities and with a good inter-grain connectivity. **b,** Temperature dependence of the normalised χ in the vicinity of the superconducting transition for field cooling (FC) and zero field cooling (ZFC) showed a $T_c$ of 33 K. An external field of 1 mT was applied for $H \parallel ab$. This temperature coincides with $T_{c,0}$. **c,** Enlarged view of $\rho(T)$ in the vicinity of the transition. The blue arrow defines the normal state resistivity $\rho_n$ below which the resistivity deviates from the linear fit to the normal state (red line).

## 5. *E-J* characteristics for evaluating $J_c$ for various fields and temperatures

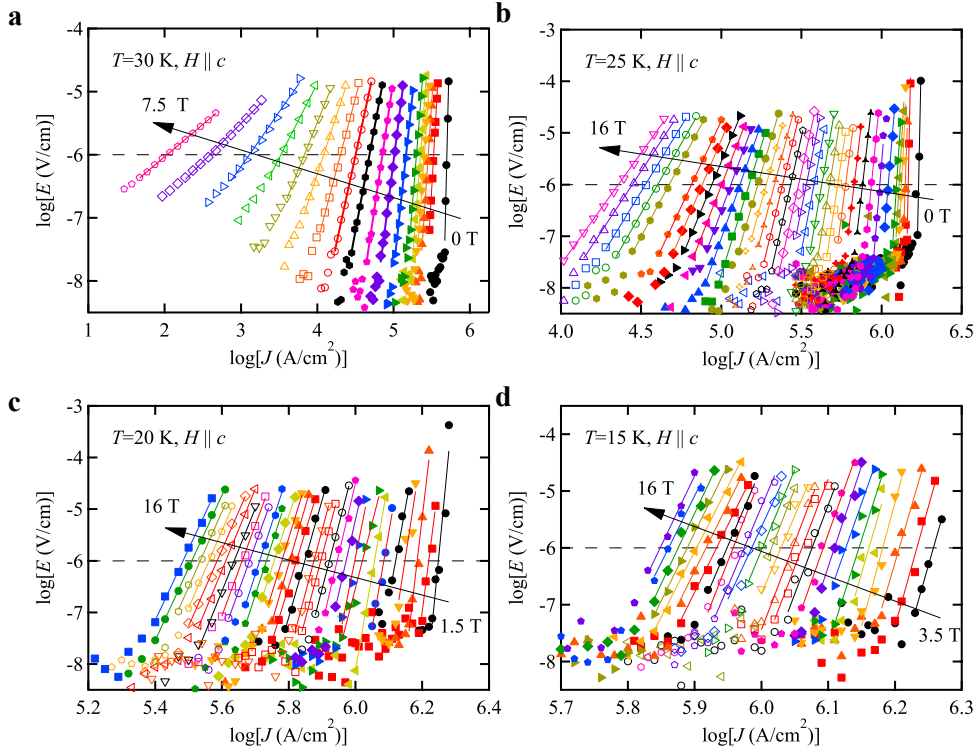

**Figure S5 | Log-log plot of *E-J* characteristics for the K-doped Ba122 thin film.  a,** *E-J* curves measured for $H \parallel c$ at *T*=30 K, **b,** 25 K, **c,** 20 K, and **d,** 15 K. The dashed line is the electric field criterion $E_c$ for determining $J_c$. All curves show a power law behaviour, indicative of the absence of weak-link.

## 6. Magnetisation measurement to evaluate $J_c$

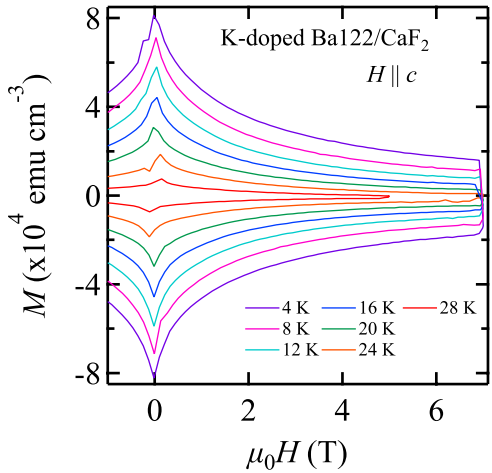

**Figure S6| Field dependence of magnetisation *M*.** *M-H* hysteresis loops measured at various temperatures. Applied magnetic field is parallel to the *c*-axis.